\documentclass[12pt]{iopart}

\newcommand{\op}[1]{\ensuremath{\hat{#1}}}
\newcommand{\mum}[1]{\ensuremath{#1\mu m}}

\newcommand{\qa}[1]{\textsf{`}#1\textsf{'}}
\renewcommand{\vec}[1]{\ensuremath{\mathbf{#1}}}

\usepackage{iopams}  
\usepackage{fancyheadings}
\usepackage[dvipdfm]{graphicx}
\begin{document}

\title[Vortex-antivortex pair dynamics in an exciton-polariton condensate]{Vortex-antivortex pair dynamics in an exciton-polariton condensate}

\author{M D Fraser$^{1,2}$, G Roumpos$^{3}$ and Y Yamamoto$^{2,3}$}

\address{$^1$ Institute for Nano Quantum Information Electronics, 
University of Tokyo, 4-6-1 Komaba, Meguro-ku, 
Tokyo 153-8505, Japan}
\address{$^2$ National Institute of Informatics, 
2-1-2 Hitotsubashi, Chiyoda-ku, 
Tokyo 101-8430, Japan}
\address{$^3$ Edward L. Ginzton Laboratory, 
Stanford University, Stanford, 
California 94305-4085, USA}

\ead{mdfraser@nii.ac.jp}

\begin{abstract}
The study of superfluid and Berezinskii-Kosterlitz-Thouless phases in exciton-polaritons requires an understanding of vortex dynamics in a dissipative unconfined condensate.  In this article we study the motion of dynamic vortex-antivortex pairs and show that vortex pair stability defined as ordered motion as opposed to rapid separation or recombination is the result of balance between dissipative velocities in the condensate and interaction with thermal polaritons. The addition of a trapping potential is further shown to considerably enhance the lifetime of a single vortex pair in this system. These investigations have important consequences for interpretation of recent results and future investigations of two-dimensional superfluid phases in polariton condensates.

\end{abstract}

\pacs{03.75.Kk, 47.32.C, 67.25.dk, 71.36.+c}

\setcounter{page}{1}
\tableofcontents
\maketitle

\pagestyle{fancy}  
\fancyhf{}  
\setlength{\headheight}{15pt}  
\lhead{}  
\rhead{\thepage}  
\renewcommand{\headrulewidth}{0pt}  
\fancypagestyle{plain}{
\fancyhf{} 
\renewcommand{\headrulewidth}{0pt}  
\renewcommand{\footrulewidth}{0pt}}

\section{Introduction: Two-dimensional polariton condensation}

Superfluid behavior is well known in liquid helium and atomic Bose-Einstein condensates and extensive studies have been made of the properties of quantized vortices in these systems \cite{aboshaeer2001ovl,donnelly1991qvh}.  A newer type of condensate receiving a lot of current attention, occurs not in a gas or liquid of \qa{real} particles, but instead occurs in quasi-particles in semiconductors known as exciton-polaritons (hereafter referred to as polaritons) \cite{deng2002csm,kasprzak2006bec}.  While the condensation of bosonic excitonic particles has been studied for many years, the strong coupling of excitons and photons reduces the effective mass by such a degree that condensation occurs in GaAs at temperatures of order 10K \cite{deng2002csm} and is expected at room temperature in other materials \cite{kavokin2003pla}.  However, the small effective mass comes at the expense of finite particle lifetime, implying that although polaritons can reach equilibrium with the lattice \cite{deng2006qde}, a steady-
state condensate is formed in place of a condensate in true thermal equilibrium.  One of the advantages of a condensate embedded within a semiconductor matrix is the ability to manipulate the internal properties and dimensionality through material changes and the use of heterostructures.  Though not strictly a necessary geometry, many of the recent observations of polariton condensation occur in two-dimensional (2D) quantum wells embedded within microcavities in order to reach the strong coupling regime.

Superfluidity is an expected property of this condensate-like system \cite{utsunomiya2008obe,amo2009collective}, and though not direct evidence of this, quantized vortices have already been observed \cite{lagoudakis2008qve,roumpos2009}.  Quantized vortices are topological excitations of phase coherent systems and exist commonly in response to an applied rotational field, but in two-dimensions spontaneous formation as a result of phase fluctuations is also possible \cite{andersen2002pfa}.  Spontaneous vortex excitation is known to destroy long range order in a reduced dimensional infinite system preventing superfluidity \cite{mermin1966afo,hohenberg1967elr} at non-zero temperature.  However, below a certain temperature, the thermal energy is not sufficient to generate distinct single vortices, but only the lower energy vortex-antivortex pairs (or vortex pairs for short).  As the vortex-induced phase gradient is now largely localized between the vortex pair, superfluidity can be recovered.  This 2D superfluid 
phase characterized by the presence of bound vortex-antivortex pairs is known as the Berezinskii-Kosterlitz-Thouless (BKT) phase, occurring at temperatures below $T_{BKT}$ \cite{kosterlitz1973oma}.

A Bose-Einstein condensate like transition can also be recovered in a 2D system providing the condensate is confined to a finite area of a size comparable to the phonon de Broglie wavelength in the fluid, thereby excluding long-wavelength fluctuations through a discrete density of states \cite{petrov2000bec}.  Although the transition from a normal (thermal) state to a BKT phase is discontinuous at $T_{BKT}$, the transition from BEC to BKT is a continuous function of condensate size and temperature, both contributing to the vortex pair density in the system \cite{simula2006tav}.  The polariton condensate is thus an ideal system in which to study crossovers between various 2D condensate and superfluid phases.

The BKT phase has previously been reported in superfluid $^{4}$He films \cite{bishop1978sst}, atomic hydrogen films \cite{safonov1998oqt} and dilute gas atomic BEC \cite{hadzibabic2006bkt}.  In each of these cases however, although convincing evidence for the BKT phase is presented, the microscopic nature of the state (the vortex-antivortex pairs) has yet to be observed.  The small characteristic size scale of vortices (healing length $\xi$) to the condensate size, results in a population of vortex pairs and/or free vortices far in excess of one, making single vortex pair observation difficult.  The evidence supporting the BKT phase in atomic BEC however, is the observation of a proliferation of free vortices with increased temperature believed to be due to pair breaking when the temperature exceeds the pair binding energy. 

Recently however, direct evidence of a single vortex-antivortex pair has been reported in an exciton-polariton condensate \cite{roumpos2009} . The observation is based mainly on phase dislocations in interferometry experiments indicative of a single vortex-antivortex pair. Due to the low polariton effective mass $m\sim10^{-5}m^{0}_{e}$ ($m^{0}_{e}$ being the free electron mass), the vortices are significantly larger and for a typical condensate density and radius ($L\sim\mum{10}-\mum{15}$) one vortex pair can be comparable to the system size.  Furthermore, unlike previous quantized vortex observations in polariton condensates, due to the low disorder in this sample ($\tilde{V}_{d}\lesssim0.1meV$), this vortex pair is believed to be unpinned.  This recent observation in particular suggests the necessity of a proper understanding of vortex and vortex pair dynamics in this system.

In contrast to the superfluid He and atomic BEC systems which generally have a particle number conservation over a dynamical timescale, the polariton condensate occurs in quasi-equilibrium, the result of a steady-state process of continuous stimulated scattering from a polariton reservoir (pump) and the finite polariton lifetime $\tau_{pol}\sim ps$ \cite{wouters2007enb}.  While the properties and dynamics of quantized vortices \cite{caradocdavies1999cdv,tsubota2002vlf} and vortex pairs \cite{crasovan2003svd} have received considerable theoretical attention in superfluid helium and atomic condensates, given its steady-state dissipative nature, the polariton condensate phase is distinct and the dynamical nature of quantized vortices presently unknown.  The theoretical investigation of vortex dynamics presented here is directly relevant to the recent experimental observation of a single vortex pair and to further investigations of vortex nucleation and BEC-BKT crossover in a polariton condensate. 

\section{Dissipative Gross-Pitaevskii equation}

Condensate dynamics are usually modeled using a form of the time dependent Gross-Pitaevskii equation (GPE) to describe evolution of the condensate order parameter $\psi(\vec{r},t)$ \cite{pitaevskiui2003bec}.  This technique has been extended to describe non-condensate components including quantum and thermal depletion and finite temperature effects \cite{giorgini1996cfa,hutchinson1998gft}.  In this work we apply a dissipative GPE previously shown to contain the essential parameters necessary for simulation of polariton condensates \cite{wouters2007enb,keeling2008srv,lagoudakis2008qve}.  This dissipative GPE is coupled to a thermal reservoir population $n_{R}(\vec{r},t)$ by stimulated scattering ($R(n_{R}(\vec{r},t))$) and interactions (with coupling constant $g_{R}$).  Our study here specifically differs from other studies of vortex pair dynamics in that the non-condensate reservoir population is not uniquely determined by coupling with the condensate, but also by the spatial pumping profile.  In addition 
because of the steady-state nature, this unconfined condensate possesses a unique velocity profile.  

We start by defining the usual single particle Hamiltonian, 

\begin{equation}
    \label{eq:singleparticleH}
    \op{H}_{0} = -\frac{\hbar^{2}\nabla^{2}}{2m}+V_{ext}(\vec{r})
\end{equation}

\noindent
where $m$ is the polariton effective mass, and $V_{ext}(\vec{r})$ is any external confining or disorder potential profile.  The dissipative time-dependent Gross-Pitaevskii equation is then given as 

\begin{equation}
    \label{eq:gped}
    \fl  i\hbar\frac{\partial\psi(\vec{r},t)}{\partial t} = \left(\op{H}_{0}-\frac{i\hbar}{2}\left[\gamma_{C}-R(n_{R}(\vec{r},t))\right]+g_{C}|\psi(\vec{r},t)|^{2} +g_{R}n_{R}(\vec{r},t)\right)\psi(\vec{r},t)
\end{equation}

\noindent
where the loss and gain terms of $-i\hbar\gamma_{C}/2$ and $i\hbar R(n_{R}(\vec{r},t))/2$ respectively describe the process of polariton decay as photons leaving the cavity at rate $\gamma_{C}$ (or alternatively with lifetime $\tau_{C}$) and stimulated scattering into the condensate at a rate $R(n_{R}(\vec{r},t))$ determined by the reservoir population distribution $n_{R}(\vec{r},t)$.  This reservoir population is described by a rate equation model, 

\begin{equation}
    \label{eq:rteq}
    \frac{\partial n_{R}(\vec{r},t)}{\partial t} = P_{l}(\vec{r}) - \gamma_{R}n_{R}(\vec{r},t) - R(n_{R}(\vec{r},t))|\psi(\vec{r},t)|^{2} 
\end{equation}

\noindent
where $P_{l}(\vec{r})$ is the laser excitation profile (in photons/$\mu m^{2}ps$) and $\gamma_{R}$ is the reservoir polariton loss rate.

Previous investigation of vortices with this model have not considered vortex pair dynamics, and it is not presently clear how a vortex pair should behave in a dissipative polariton condensate.  Previous experimental observations of quantized vortices in polariton condensates have been carried out in materials with a level of disorder potential sufficient to spatially pin the vortex, prohibiting any vortex dynamics and allowing simple experimental observation in time-integrated measurements.  However, the recent clear observation of a spontaneously formed vortex-antivortex pair occurs in the GaAs-based system where the disorder is low enough for the vortex pair to remain unpinned and as such, dynamics and stability are critical to understanding this experimental result and is the main motivation for this research.

\section{Dissipative vortex pair dynamics}

The key parameters in the model of equations \ref{eq:gped} and \ref{eq:rteq} which are expected to influence vortex dynamics in the absence of any trapping potential are the condensate density $n_{C}$, reservoir density $n_{R}$, scattering rate $R(n_{R}(\vec{r},t))$ and condensate polariton lifetime $\tau_{C}$.  It is difficult to study the effect of any one of these parameter independently, as in this dynamic model changing either the lifetime or scattering rate both alter the final steady state condensate population.  As the condensate density critically determines the vortex size through the healing length parameter $\xi=\hbar/\sqrt{2mg_{C}n_{C}}$, both $R(n_{R}(\vec{r},t))$ and $\tau_{C}$ are adjusted together in order to maintain an approximately constant steady state condensate density.  Equation \ref{eq:steadystateR} describes the steady state relationship between these parameters (for a uniform system).

\begin{equation}
    \label{eq:steadystateR}
    \frac{R(n_{R}(\vec{r}))}{n_{R}(\vec{r})} = \frac{\gamma_{R}}{P_{l}/\gamma_{C} -n_{C}}
\end{equation}

\begin{figure}[!h]
\centering
\includegraphics[width=0.7\linewidth]{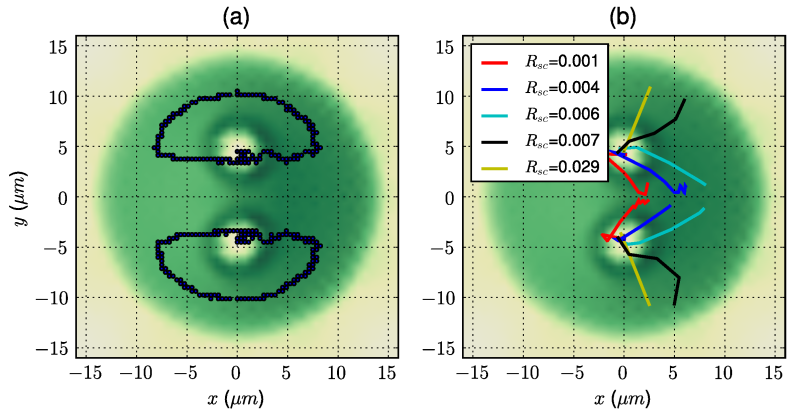}
\caption{\baselineskip=0.77\baselineskip
The vortex pair trajectory for (a) a conservative condensate in a square well (abrupt walls of $1meV$ depth) circularly symmetric trapping potential of radius $r = \mum{15}$, and (b) dissipative condensates at various points in the parameter space depicting a range of paths between rapid separation and rapid recombination.  The corresponding scattering rate coefficient is indicated (in $\mu m^{2}ps^{-1}$).  The initial vortex pair positions are $r_{v}=\pm\mum{4.5}$ along the $y$-axis and the pumping spot radius is $L = \mum{15}$.}
\label{fig:trajectory}
\end{figure}

\noindent
The scattering rate $R(n_{R}(\vec{r}))$ is given a linear dependence on $n_{R}(\vec{r})$ with coefficient $R_{sc}$.  The study of dissipative vortex-antivortex pair dynamics here specifically addresses GaAs-based polariton condensates, however general arguments will be drawn relevant to all other material systems.  From a manually ascribed initial state, the condensate is allowed to evolve in time until a steady state distribution is achieved (constant energy and static particle distribution).  The vortex pair phase profiles are then artificially imprinted along the $y$-axis (at positions $\pm d_{v}/2$ where $d_{v}$ is the vortex pair separation), with a mirror symmetry about the $x$-axis.  The dynamics of an unpinned vortex-antivortex pair in a spatially-infinite conservative (non-dissipative) superfluid is simple and well understood.  Due to the interaction of the mutual phase gradients the vortex-antivortex pair undergoes a linear motion perpendicular to its dipole axis, with direction determined by the 
dipole orientation \cite{donnelly1991qvh} (here, along the $x$-axis).  The vortex pair will then travel at a velocity dependent on the vortex separation $d_{v}$ given by equation \ref{eq:vortexpairvelocity}. 

\begin{equation}
    \label{eq:vortexpairvelocity}
    |\vec{v}_{p}|=\frac{{|\kappa|}}{2\pi d_{v}}   
\end{equation}

\noindent 
where $\kappa=l(\hbar/m)\hat{z}$ is the quantized circulation of each vortex with vector perpendicular to the $x$-$y$ plane and $l=\pm 1$.  In the absence of any dissipation, the vortex pair will maintain separation of $d_{v}$.  The introduction of a confining potential $V_{ext}(\vec{r})$ creates a boundary with which the vortex pair interacts, perturbing its one-dimensional motion.  The form of the perturbation depends on the type of confining potential. In a conservative system, the vortex pair will track the boundary to preserve the energy of the vortex pair periodically returning to its starting position.  Figure \ref{fig:trajectory}a shows the numerical solution of the conservative GPE in a square well trap (abrupt walls) giving an example of this type of motion.  The trajectory in a harmonic trap is qualitatively similar, but this example is closer to that of a top-hat pumping profile considered in the rest of this paper for the dissipative condensate.

\subsection{Classification of vortex pair trajectories}

The vortex pair dynamics in a dissipative system are found to differ considerably.  The initially imprinted pair with separation $d_{v}$ is found instead to choose from a continuum of different trajectories based on the parameters chosen for the simulation, some examples of which are shown in figure \ref{fig:trajectory}b.  These motions range from the vortex pair splitting and leaving the condensate directly to recombining rapidly long before reaching the edge of the condensate.  In these simulations, a condensate density $n_{C}\approx\mum{570}^{-2}$ and a pumping spot of radius $L = \mum{15}$ are used.  This condensate size is chosen such that the vortex pair is not initially perturbed by the presence of a boundary and the motion largely independent of the condensate size can be initially observed, from which general conclusions are drawn.

\begin{figure}[!h]
\centering
\includegraphics[width=0.7\linewidth]{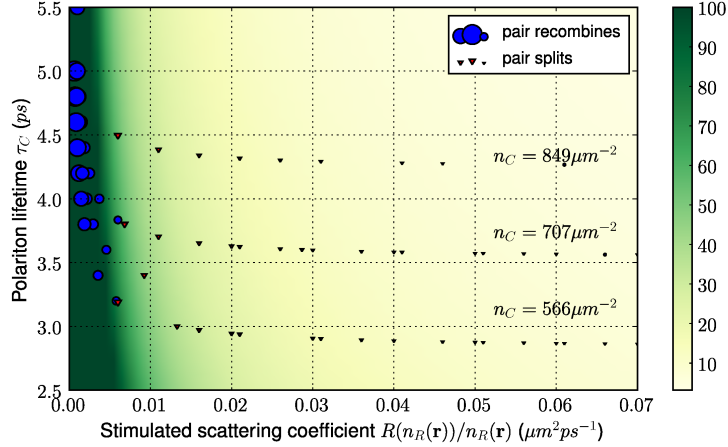}
\caption{\baselineskip=0.77\baselineskip
Map of $n_{R}$ (depicted as the circle/triangle marker size) with scattering rate $R(n_{R}(\vec{r}))$ and condensate polariton lifetime $\tau_{C}$.  The condensate density is held relatively constant in these simulations and corresponds to the different lines labeled with $n_{C}$, but does not have a significant effect on the vortex trajectory.  The vortices that leave the condensate are indicated as red triangles and those that recombine as blue circles.    The background in this plot comprises the approximate steady-state analytical form of $n_{R}$, with magnitude indicated by the colour bar in $\mu m^{-2}$.}
\label{fig:vortexclassification}
\end{figure}

To gain an understanding of the effect of these parameters, a map of $n_{R}$ (size of circles) with variation in $R(n_{R}(\vec{r},t))/n_{R}(\vec{r},t)$ and $\tau_{C}$ is plotted figure \ref{fig:vortexclassification}.  The background map (saturated at small scattering rate) is the steady-state analytical contribution (equation \ref{eq:steadystateR}).  In this parameter space the vortex pair trajectories are divided into two groups according to whether the vortex pair recombines in the center of the condensate before reaching the edge (depicted as blue circles) and vortex pairs splitting and leaving the condensate (red triangles).  A clear transition between these two different groupings is apparent and observed to be largely dependent on $R(n_{R}(\vec{r},t))$.

If the initial vortex pair spacing $d_{v}$ is altered in this simulation, it does not affect this trend other than to slightly shift the transition between the two classifications of vortex pair evolution.  This remains the case provided the vortex pairs are not initially overlapping slightly ($d_{v}\lesssim2\xi$) or the vortex is initially separated from the boundary by an amount of order $|L-r_{v}|\lesssim2\xi$.  In long timescale experimental scenarios where a statistical ensemble of initial vortex pair separations is expected this then corresponds to the same trend but with a blurred cross-over transition between the two classifications.  The condensate population is also found to not impose much effect on this crossover (within experimentally reasonable ranges) as indicated in figure \ref{fig:vortexclassification} where the different lines indicate different condensate population density.

\begin{figure}[!h]
\centering
\includegraphics[width=0.7\linewidth]{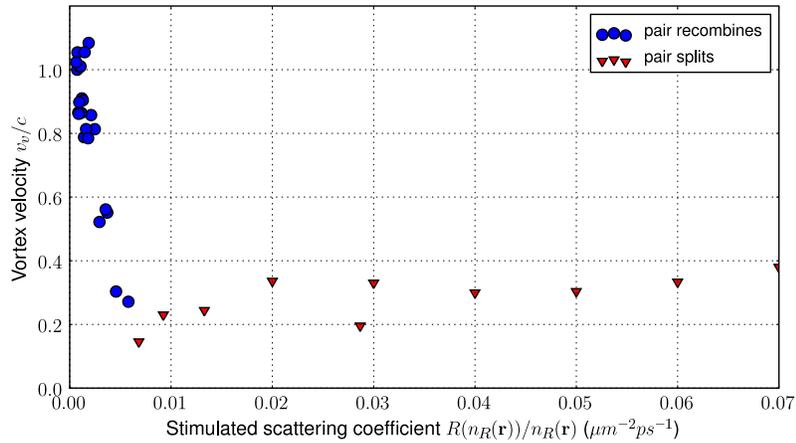}
\caption{\baselineskip=0.77\baselineskip
A map of the numerical vortex velocities as a function of the $R(n_{R}(\vec{r}))$ and $\tau_{C}$ parameter space.  The recombining vortex pairs are again indicated as blue circles, and the radially separating ones as red triangles.}
\label{fig:vortexvelocity}
\end{figure}

Attention is now turned to a map of vortex velocities constructed from this numerical data and shown in figure \ref{fig:vortexvelocity}.  The vortex velocities are presented as a fraction of the condensate sound velocity defined as $c = \sqrt{g_{C}n_{C}/m}$.  The distinction between the two classifications of vortex dynamics is further clear in this data.  For very low scattering rates (which corresponds to high reservoir population density), the vortex pair recombines rapidly.  This vortex velocity drops rapidly from the sound velocity as the scattering rate is increased, stopping when the cross-over to radially-separating vortex pairs is achieved, remaining roughly constant thereafter at a fraction of the sound velocity.  This analysis suggests two different mechanisms altering the vortex pair velocity vector.

\subsection{Contributions to vortex velocity}

In a conservative condensate, the absence of any forces implies the vortex velocity $\vec{v}_{L}$ will coincide with that of the local superfluid flow $\vec{v}_{L} = \vec{v}_{s}$ in correspondence with the familiar Magnus force (equation \ref{eq:magnus}) \cite{sonin1997mfs,donnelly1991qvh}.  In rotating trapped condensate for example, the vortex lattice will rotate at the same angular velocity as the condensate and thus appear stationary in the rotating frame.  In a dissipative system, i.e. through interaction with thermal population and excitations at finite temperature, energy is transferred between these non-condensate populations and energetically unfavourable vortex states will decay by approaching the condensate boundary decaying via excitations at the edge \cite{fedichev1999ddv}.

\begin{equation}
    \label{eq:magnus}
    \vec{f}_{M} = n_{C}\vec{\kappa}\times(\vec{v}_{L} - \vec{v}_{s})
\end{equation}

The superfluid velocity $\vec{v}_{s}$ is defined as being the superfluid velocity far from the vortex line.  Thus, in the present vortex pair system, the superfluid velocity $\vec{v}_{s}$ consists of velocity contributions from each of the two vortices ($\vec{v}_{L1}$ and $\vec{v}_{L2}$) and the radially dissipating polaritons due to repulsive interactions and a lack of confinement.  Neglecting the condensate boundary, to calculate the local velocity field at the vortex $v1$ (at $\vec{r}_{v1}$ with $\vec{\kappa}_{v1}$) in the presence of vortex $v2$ (at $\vec{r}_{v2}$ with $\vec{\kappa}_{v2}$) and the radial velocity gradient ($\vec{v}_{C}(\vec{r})$) we can use, 

\begin{equation}
    \label{eq:condvelocity_vs}
    \vec{v}_{s}(\vec{r}_{v1}) = \frac{\vec{\kappa_{v2}}\times(\vec{r}_{v1}-\vec{r}_{v2})}{|\vec{r}_{v1}-\vec{r}_{v2}|^{2}} + \vec{v}_{C}(\vec{r})
\end{equation}

The second source of forces on the vortex pair is the interaction of vortices with non-condensate population which creates a drag force dependent on the interaction energy with this non-condensate population $n_{R}$.  This force is commonly broken up into longitudinal and transverse components,

\begin{equation}
    \label{eq:dragforce}
    \vec{f}_{D} = -n_{C}\frac{Bn_{R}}{2(n_{C}+n_{R})} \kappa\times\left[\frac{\kappa}{|\kappa|}\times(\vec{v}_{n}-\vec{v}_{s})\right]-n_{C}\frac{B'n_{R}}{2(n_{C}+n_{R})} \kappa\times(\vec{v}_{n}-\vec{v}_{s})
\end{equation}

\noindent
where $\vec{v}_{n}$ is the velocity of the normal component.  The origin of these drag forces is usually attributed to interaction with thermally excited non-condensate modes, where the perpendicular component is commonly known as the Iordanskii force \cite{geller1998transverse,thouless2001vortex}.  The form of the friction coefficients $B$ and $B'$ have been evaluated theoretically \cite{sonin1997mfs,fedichev1999ddv} and experimentally \cite{donnelly1991qvh}, though exact determination is very dependent on the exact details of the system and the contributions to the magnitude of the Iordanskii force are particularly complex.  We find however that conclusions can be drawn relatively independent of the exact magnitude of these coefficients.  

\subsection{Condensate velocity profiles}

\begin{figure}[!h]
\centering
\includegraphics[width=0.7\linewidth]{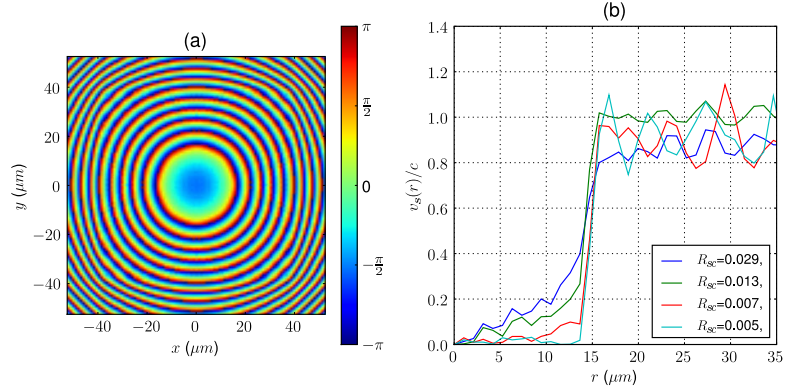}
\caption{\baselineskip=0.77\baselineskip
(a) Steady-state unconfined condensate phase profile $S(\vec{r})$ and (b) the radial cross-section of the velocity profile.  The particle velocity within the condensate increases with the scattering coefficient $R(n_{R}(\vec{r}))/n_{R}(\vec{r})$ in response to a reduction in drag forces through decreasing reservoir density $n_{R}(\vec{r})$}
\label{fig:condensatevelocity}
\end{figure}

As the condensate is unconfined, we expect there to be some radially dissipative condensate velocity in the system amounting to a continual radial loss of particles due to particle repulsive interactions, giving the form of $\vec{v}_{C}(\vec{r})$.  Figure \ref{fig:condensatevelocity}a shows the numerically-evaluated steady-state unconfined condensate phase profile $S(\vec{r})$.  The condensate velocity is simply the gradient of the phase ($\vec{v}_{s}=\hbar/m \nabla S(\vec{r})$) and the radial velocity profile can be calculated as in figure \ref{fig:condensatevelocity}b.  The point at which the velocity increases discontinuously is the approximate condensate boundary.  Note that the velocities within the condensate possess a similar fraction of $c$ as do those determined for splitting pairs in figure \ref{fig:vortexvelocity}.

In the absence of energy dissipation, we expect vortices in the condensate to move with the same velocity vector as the local superfluid flow.  The contributions to the magnitude of the vector $v_{C}(\vec{r})$ are from the same contributions as for the condensate sound velocity, namely, the condensate interaction energy $g_{C}|\psi(\vec{r})|^{2}$ and the effective mass $m$.  The condensate radial velocity also exhibits an inverse correspondence with the reservoir density $n_{R}(\vec{r})$ (induced by a reduction in scattering coefficient $R_{sc}$).  The presence of this thermal reservoir with which the condensate interacts introduces a drag force on the vortex pair.  As the energy of a vortex pair is proportional to its separation (equation \ref{eq:vortexpairenergy}), in the presence of finite $g_{R}n_{R}$, energy will be transferred from the vortex pair to the thermal reservoir, inducing a drag force on each vortex towards the pair midpoint.

\begin{equation}
    \label{eq:vortexpairenergy}
    E_{v}=\frac{n_{C}|\kappa|^{2}}{2\pi}\ln\left(\frac{d_{v}}{\xi} \right)
\end{equation} 

Based on this simple analysis, the occurrence of two distinct modes of vortex motion differing from that of a conservative condensate appear to arise from contributions of two separate sources, namely the dissipative velocity of an unconfined condensate and the drag induced by interaction with thermal particles.  Allowing the two forces to cancel ($\vec{f}_{M}+\vec{f}_{D}=0$), the velocity vector $\vec{v}_{L1}$ of vortex $v1$ is given by,

\begin{equation}
    \label{eq:vortexpairvL1}
    \fl  \vec{v}_{L1} = \vec{v}_{s} + \left\lbrace \frac{Bn_{R}|\kappa_{v1}|}{2(n_{C}+n_{R})} \hat{\kappa}_{v1}\times (\vec{v}_{n}-\vec{v}_{s}) - \frac{B'n_{R}|\kappa_{v1}|}{2(n_{C}+n_{R})} \hat{\kappa}_{v1}\times \left[\hat{\kappa}_{v1}\times(\vec{v}_{n}-\vec{v}_{s}) \right] \right\rbrace
\end{equation} 

\noindent
where $\hat{\kappa}_{v1} = \vec{\kappa}_{v1}/|\kappa_{v1}|$.  Making substitutions for $\vec{v}_{s}$ relevant for this specific case (see equation \ref{eq:condvelocity_vs}) reveals the contributions of these two sources to $\vec{v}_{L1}$.  If we assume the two vortices are initially at positions $\vec{r}_{v1} = (d_{v}/2)\hat{y}$ and $\vec{r}_{v2} = -(d_{v}/2)\hat{y}$ with circulation vectors of $\vec{\kappa}_{v1}=+(\hbar/m)\hat{z}$ and $\vec{\kappa}_{v2}=-(\hbar/m)\hat{z}$ respectively yields the form of $v_{s}$ for $v1$,

\begin{equation}
    \label{eq:vortexpairvs1}
    \vec{v}_{s}(\vec{r}_{v1}) = +\frac{\hbar}{md_{v}} \hat{x} + |\vec{v}_{C}(d_{v}/2)|\hat{y} 
\end{equation} 

\begin{figure}[!h]
\centering
\includegraphics[width=0.7\linewidth]{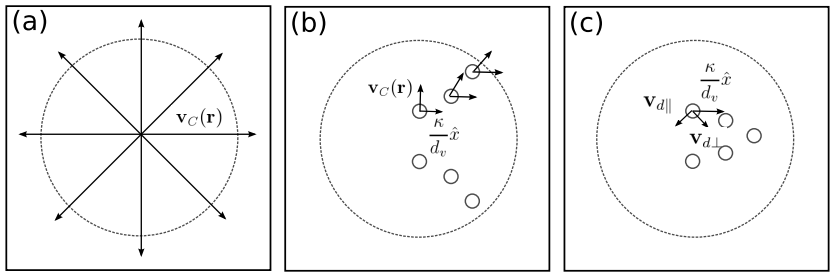}
\caption{\baselineskip=0.77\baselineskip
Schematic illustration of the different components contributing to the vortex velocity, where the large dotted circle indicates the approximate condensate boundary and the small circles represent a sample of instantaneous vortex positions with time increasing to the right of the schematic.  (a) shows the radial dissipative velocity $\vec{v}_{C}(\vec{r})$, the result of repulsive interactions and lack of confinement.  When the velocity in (a) is dominant, the vortex vector in (b) results and the vortex pair separates.  When drag forces dominate due to large $n_{R}$, the vortex pair recombines as in (c).}
\label{fig:vortexvelocityschem}
\end{figure}

Figure \ref{fig:vortexvelocityschem} shows a schematic of various contributions to the vortex velocity with (a) showing $\vec{v}_{C}(\vec{r})$ and (b) and (c) showing the radially diverging and recombining vortex pairs respectively.  The first term in equation \ref{eq:vortexpairvL1} corresponds to the contribution of dissipative superfluid flow and the second term describing contributions of non-condensate interaction effects.  While the exact trajectory depends on the accurate coefficients, it is clear that the first term contributes an outward velocity in the $+\hat{y}$-direction with some curvature due to the $+\hat{x}$ component as a result of the vortex pair linear trajectory.  Thus, on its own (i.e. when $n_{R}$ is small compared to $n_{C}$), it describes a splitting and radially dissipating vortex pair as depicted in figure \ref{fig:vortexvelocityschem}b.  If we ignore the presence of the radial dissipative velocity ($v_{C}(\vec{r})=0$), the effect of the drag forces on $v1$ in the second term of 
equation \ref{eq:vortexpairvL1} can be described, which are proportional to $n_{R}$.  The perpendicular component labeled $\vec{v}_{d\perp}=(Bn_{R}/2(n_{C}+n_{R})) \hat{\kappa}\times (\vec{v}_{n}-\vec{v}_{s})$ contains components in directions $+\hat{x}$ and $-\hat{y}$ while the longitudinal drag $\vec{v}_{d\parallel}=-(B'n_{R}/2(n_{C}+n_{R}))\hat{\kappa}\times [\hat{\kappa}\times(\vec{v}_{n}-\vec{v}_{s})]$ contains vector components in directions $-\hat{x}$ and $-\hat{y}$ as illustrated in \ref{fig:vortexvelocityschem}c.  Thus, independent of the magnitude of the coefficients $B$ and $B'$ this set of equations generally describes a trajectory directed inwards ($-\hat{y}$-direction) towards the mid point of the vortex pair.  Clearly, if this analysis is performed again with the second vortex $v2$ the same results will be achieved with oppositely directed $\hat{y}$-vectors, demonstrating that this simple analytical model agrees with our previous numerical results.

\subsection{Effect of trapping potential and interactions on vortex pair motion}

The application of a trapping potential will prevent radial particle escape by limiting any radial superflow present in the system and its presence can be used to check the effect of this superflow on vortex pair motion.  Furthermore, the onset of the BKT-phase occurs not only with temperature, but also with confinement size.  In most current experiments the finite laser pumping area is sufficient to restrict the condensate area (due to finite lifetime and diffusion length). Experimental investigations of BKT transitions in a polariton condensate are likely to benefit from the use of a trapping potential to confine the gas, particularly if the lifetime is increased.  Effective square well and harmonic profile traps for polaritons have been previously experimentally demonstrated \cite{lai2007czs,utsunomiya2008obe,balili2007bec} allowing greater control over the condensate density profile.  

\begin{figure}[!h]
\centering
\includegraphics[width=0.7\linewidth]{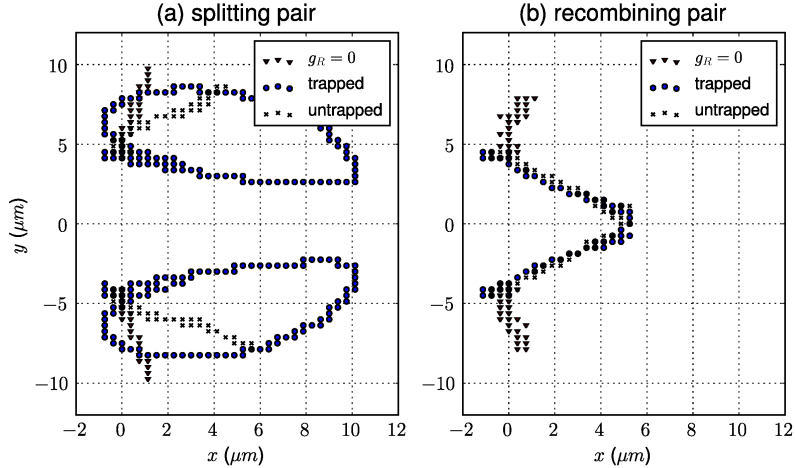}
\caption{\baselineskip=0.77\baselineskip
Vortex pair trajectories for parameters yielding (a) radially diverging vortex pair and (b) recombining vortex pair.  Within these plots the effects of applying a trapping potential and turning off the interaction with the reservoir component are also displayed as separate trajectories.}
\label{fig:trappingandinteractions}
\end{figure}

In figure \ref{fig:trappingandinteractions} we show the numerical trajectories for (a) a diverging vortex pair and (b) for a recombining vortex pair.  In figure \ref{fig:trappingandinteractions}a, when a trapping potential is added, the vortex pair does not immediately split and leave the condensate, but recovers some of its circular motion remnant of the conservative scenario.  This implies that the velocity of the radial dissipative flow as expected is the main contributor to perturbing the motion for this parameter range.  In this case, the presence of a trapping potential extends the vortex pair lifetime by many times.  In figure \ref{fig:trappingandinteractions}b where the parameters dictate a rapidly recombining vortex pair, adding a trapping potential has no obvious effect on the trajectory, implying influence of any the radial condensate velocity is negligible for these parameters.

Figure \ref{fig:trappingandinteractions} also demonstrates the effect of turning off the interactions with the reservoir polaritons (setting $g_{R}=0$).  In figure \ref{fig:trappingandinteractions}a the effect of turning off interactions only is to reduce drag forces towards the condensate centre and allows the vortex pair to leave the condensate more directly, via a shorter path.  However, in figure \ref{fig:trappingandinteractions}b for the recombining pair, turning off reservoir interactions completely prevents the recombination of the vortex pair and the vortices leave the condensate directly with the velocity vector comparable to $v_{s}$.

\section{Conclusion}

We have studied the dynamics and stability of a single vortex pair in a dissipative model of a polariton condensate.  While vortex pairs are essentially stable against recombination and radial dissipation in conservative condensates, in an unconfined dissipative condensate, the vortex pair either splits and leaves the condensate or recombines quickly a short distance from the nucleation location. It is found that the cross-over of these two behaviours is a result of competition between a radially outward force due to the radially dissipating condensate polaritons and the interaction of the vortices with non-condensate population which strongly inhibits vortex motion and induces a drag force towards the pair midpoint as the vortex pair loses energy to the reservoir.  We note that the long time-scale vortex pair stability appropriate to conservative (atomic) condensates can be recovered in this system through the application of a trapping potential and pumping the system such that the reservoir density is not 
excessive.  These observations thus have direct relevance to the interpretation of recent observations of a dynamic vortex-antivortex pair in a polariton condensate and to the extension of these studies to push the system into the BKT regime.

\ack
This work is supported by the Special Coordination Funds for Promoting Science and Technology, Navy/SPAWAR Grant N66001-09-1-2024, and MEXT.

\end{document}